\documentclass[12pt]{article}
\newtheorem{theorem}{Theorem}
\newtheorem{lemma}{Lemma}
\newtheorem{proposition}{Proposition}

\usepackage{latexsym}

\begin{document}

\title{On the area of the symmetry orbits in $T^2$ symmetric spacetimes}
\author{James Isenberg\\
Department of Mathematics\\ and\\ Institute of Theoretical Science\\
University of Oregon\\Eugene, Oregon, USA 97403\\
jim@newton.uoregon.edu\\
\\Marsha Weaver\\Theoretical Physics Institute\\
University of Alberta\\Edmonton, Alberta, Canada T6G 2J1\\
mweaver@phys.ualberta.ca}
\date{}

\maketitle

\begin{abstract}
We obtain a global existence result for the Einstein equations.
We show that in the maximal Cauchy development
of vacuum $T^2$ symmetric initial data with nonvanishing
twist constant, except for the special case of flat Kasner
initial data, the area of the $T^2$ group orbits takes on
all positive values.  This result shows that the areal time
coordinate $R$ which covers these spacetimes runs from zero
to infinity, with the singularity occurring at $R=0$.
\end{abstract}

\newpage

\section{Introduction}
\label{intro}
Areal coordinates provide a useful, geometrically-based, choice of
foliation and coordinate system for families of globally hyperbolic
solutions of the Einstein equations which have a closed (compact
without boundary) Cauchy surface and a two-dimensional spatially
acting isometry group.  The key feature of these coordinates
is that each of the constant time hypersurfaces (labeled by a number
``$R$'') consists of all spacetime points which lie on two-dimensional
symmetry group orbits of fixed area $R$.  The function $R$ serves as
the time coordinate for those spacetimes which admit areal coordinates.

Berger, Chru\'sciel, Isenberg, and Moncrief (``BCIM'') \cite{b1} have
shown that the maximal Cauchy development of vacuum $T^2$ symmetric
initial data with nonvanishing twist constant admits a covering
by areal coordinates.  That spacetimes of this sort with vanishing
twist constants (the Gowdy spacetimes \cite{Gow}) generically admit
a covering by areal coordinates was shown in \cite{Moncrief} and
\cite{Chr}.  For Gowdy spacetimes with spatial topology $T^3$, the
time coordinate $R$ covers the range $(0,+\infty)$ \cite{Moncrief}.
For those spacetimes with nonvanishing twist constant, BCIM show that
$R$ takes on all values in the interval $(R_0, +\infty)$ with $R_0$
some nonnegative number.  Their work does not determine which
spacetimes have $R_0=0$ and which do not.

The class of spacetimes which admit global areal coordinates has
been extended beyond the vacuum case to include the maximal
Cauchy development of $T^2$ symmetric initial data for the
Einstein-Vlasov equations with spatial topology $T^3$
(excluding the special case of $T^3 \times R^1$ Minkowski
spacetime\footnote{The area of the group orbits is constant,
so cannot serve as a time coordinate.
Also excluded are flat Kasner spacetimes in which
one chooses the $T^2$ isometry group in such a way that the area of
the group orbits is constant.}) \cite{a1,ARW}.
As in the vacuum case, for spacetimes
in this family, the orbit area time coordinate $R$ has been shown
to range from $R_0$ to $+\infty$, with no way described in
\cite{a1,ARW} to determine for which spacetimes one has $R_0=0$.
The same holds true for spacetimes with Gowdy symmetry and a
nonvanishing magnetic field orthogonal to the group orbits \cite{Weav}.

In this work, we resolve the issue of when $R_0$ vanishes for the
vacuum $T^2$ symmetric spacetimes considered in BCIM.  More
specifically, we consider the maximal Cauchy development of vacuum
$T^2$ symmetric initial data for the Einstein equations in the case
of nonvanishing twist constant and show that $R_0 = 0$ unless the
spacetime is flat Kasner.  We also note that indeed flat Kasner
initial data results in $R_0 > 0$.  We suspect that a result
similar to this one for vacuum spacetimes holds for
Einstein-Vlasov solutions, but this issue has not yet been resolved.
On the other hand, it is straightforward to adapt our proof to the
spacetimes with magnetic field mentioned in the previous paragraph.

The bulk of this paper is devoted to proving the $R_0 = 0$ result
for vacuum $T^2$ symmetric
spacetimes.  The key to the proof is the analysis of a
particular energy-like quantity ${\cal E}(R)$.  We define
${\cal E}(R)$ in section~\ref{prelim}, after using the results of
BCIM to set up areal coordinates for the spacetimes of interest,
and writing out (in section~\ref{prelim}) the vacuum Einstein field
equations in terms of these coordinates.  It is relatively
straightforward to show (see proposition~\ref{flatkasner} in
section~\ref{flatkasner}) that, presuming nonzero twist
constant, ${\cal E}(R)$ vanishes at some time $R_i$ if and only if
the spacetime generated from initial data at $R_i$ is flat Kasner.
One also readily shows (see proposition~\ref{E=0R>0}) that if
${\cal E}(R_i) = 0$ then $R_0 > 0$.  Analyzing what happens in
spacetimes with ${\cal E}(R) \neq 0$ requires more work.  We do
this in section~\ref{result}.  There, we prove via a succession of
lemmas that if a smooth spacetime has areal coordinates with the
time coordinate ranging over the interval $(R_s, +\infty)$, and if
for some $R_i \in (R_s, +\infty)$ one has ${\cal E}(R_i) \neq 0$,
then the metric functions and certain of their derivatives are
bounded.  It then follows via standard arguments that the
spacetime further extends to the areal time interval $(0,+\infty)$.
This establishes our desired result, which we formally state in
theorem~\ref{main} of section~\ref{conclusion}.  We make
some concluding remarks, and also note that there is a crushing
singularity at $R_0$, in section~\ref{conclusion}.

\section{Areal coordinates and some preliminary results}
\label{prelim}

The family of spacetimes which we study in this paper is characterized
by the following: Each member of the family is a maximally extended,
globally hyperbolic spacetime region which satisfies the vacuum
Einstein equations, which has an effectively acting $T^2$ isometry
group whose generating Killing fields are everywhere spacelike and have
nonvanishing twist quantities, and which has a compact Cauchy surface
which is invariant under the $T^2$ action.  We note that this is
exactly the family of solutions considered in BCIM \cite{b1}.  We also
note that if we were instead to assume that both quantities associated
with the twists of the Killing fields vanish, then (after excluding
the special cases mentioned in the introduction, which are
characterized by the vanishing of the
spacetime gradient of $R$ on the Cauchy surface)
we would have the Gowdy spacetimes \cite{Gow}.
If either or both of the twist quantities does not
vanish, the exceptions just mentioned are automatically excluded
(see pp 113-114 of \cite{Chr}) and
the topology of the Cauchy surface is necessarily $T^3$ \cite{Chr}.
It was shown in earlier work \cite{Chr} that each vacuum $T^2$
symmetric solution with spatial topology $T^3$ (except for the flat 
solutions with group orbits of constant area, as just mentioned)
locally admits areal coordinates $(\theta, x, y, R)$, in
terms of which the metric takes the form\footnote{The form of the
metric~(\ref{metric}) written here differs slightly from the form
used in BCIM \cite{b1} in the following ways:
1) We replace ``$t$'' by ``$R$''.  2) We have set the constant
$\lambda$ to one; this can be done without loss of generality by
rescaling $\alpha$, $e^{2U}$, $e^{2\nu}$, $R$ and $K$.  3) The area
of each group orbit in the hypersurface labeled by $R$ is exactly
$R$ rather than proportional to $R$; rescaling coordinates and
metric functions allows one to do this without loss of generality.
4) The metric ``shift" components $M_1$ and $M_2$ appearing in BCIM
have, without loss of generality, been set to zero. 5) The metric
functions $G_1$ and $G_2$ from BCIM are replaced by $G$ and $H$.}
\begin{eqnarray}
g &=& e^{2(\nu-U)}(-\alpha \,
dR^2+d\theta^2)+e^{2U}[dx+A \, dy+(G+AH) \, d\theta]^2\nonumber\\
& & +e^{-2U}R^2(dy+H \, d\theta)^2.
\label{metric}
\end{eqnarray}
Here the time coordinate $R$ locally labels spatial hypersurfaces of
the spacetime, and each such hypersurface consists of all of the $T^2$
group orbits which have area $R$.  The spatial coordinate $\theta$
is periodic, with period one.  In accord with the symmetry, the smooth
metric functions $\alpha$, $\nu$, $U$, $A$, $G$, and $H$ are all
independent of $x$ and $y$, and periodic in $\theta$.

Before we write out the Einstein equations in terms of areal
coordinates and the metric functions appearing in (\ref{metric}),
we wish to define a pair of functions which play an important role in
our analysis below.  We define $\beta$ via\footnote{Note that $\beta$
is the same function as $-\tilde{\nu}$, appearing in BCIM.}
\begin{equation}
\beta = \nu + \frac{\ln \alpha}{2};
\end{equation}
it is used as an effective replacement for the function $\nu$.  Then
we define $h$ by
\begin{equation}
h=\frac{U_R^2}{\sqrt{\alpha}}+\sqrt{\alpha} \,
U_\theta^2+\frac{e^{4U}}{4R^2}\left(\frac{A_R^2}
{\sqrt{\alpha}} +\sqrt{\alpha} A_\theta^2\right);
\label{h}
\end{equation}
note that here and below, we use the notation $U_R=\partial_R U$,
{\it etc}. As we shall see, the function $h$ plays the role of an
effective energy density for these spacetimes, in that its integral
over space is monotone and bounded in time, so long as $R>0$.

We also wish to carry out a useful simplification involving the
twist quantities. For a spacetime with a spatially acting $T^2$
isometry group generated by a pair of (commuting) Killing fields
$X$ and $Y$, the two twist quantities are given by 
\begin{equation}
\epsilon_{abcd} X^a Y^b \nabla^c X^d,
\label{vanishingtwistconstant}
\end{equation}
and 
\begin{equation}
\epsilon_{abcd} Y^a X^b \nabla^c Y^d.
\label{nonvanishingtwistconstant}
\end{equation}
If the spacetime satisfies the vacuum Einstein equations, then
both of these quantities are constant.  Now if one or both of
these twist constants is nonvanishing, then it is easily verified
that one can always choose a new set of commuting Killing fields
(constant linear combinations of $X$ and $Y$) so that the twist
quantity in expression~(\ref{vanishingtwistconstant}) vanishes,
while the other does not.  Further, one can always choose
coordinates so that one of the new Killing fields is
$\partial_x$ and the other is $\partial_y$, and so that the
metric remains in the form (\ref{metric}). This choice, which
we shall assume throughout the rest of the paper, slightly
simplifies the field equations, and also results in the
condition ${\cal E} = 0$ holding if and only if the spacetime
is flat Kasner. Note that if $X = \partial_x$
and $Y=\partial_y$ were chosen so that the twist constant
(\ref{vanishingtwistconstant}) were nonvanishing, then the
condition ${\cal E} = 0$ would not be preserved by the evolution,
and flat Kasner would have ${\cal E} \neq 0$. Thus, in the
discussion of the proof of our results, whenever the particular
form of the evolution equations is involved, or whenever the
quantity ${\cal E}$ is under consideration, the phrase ``with
nonvanishing twist constant'' should be taken to mean just
one nonvanishing twist constant, as just described. On the
other hand, the choice of Killing fields and coordinates
does not affect the area function $R$.  Thus in the statement
of the theorem and also in the abstract, the phrase ``with
nonvanishing twist constant'' should be taken to mean that
both or either one of the twist constants is nonvanishing.
This is true for other statements and results which involve $R$ only.

The vacuum Einstein equations for the metric~(\ref{metric}),
with coordinates $x$ and $y$ chosen as described in the previous
paragraph, split into three sets (see section 2 in BCIM) as follows:
\begin{description}
\item[Constraint Equations:]
\end {description}
\begin{eqnarray}
\beta_R &=& \sqrt{\alpha} R h - \frac{e^{2\beta}K^2}{4R^3},
\label{constr1}\\
\beta_\theta &=& 2R\left(U_R U_\theta + \frac{e^{4U}}{4R^2}
A_R A_\theta \right),
\label{constr2}\\
\alpha_R &=& -\frac{\alpha e^{2\beta} K^2}{R^3}.
\label{constr3}
\end{eqnarray}
\begin{description}
\item[Evolution Equations:]
\end{description}
\begin{eqnarray}
U_{RR} - \alpha U_{\theta\theta} &=& - \frac{U_R}{R}
+ \frac{\alpha_R U_R}{2\alpha} + \frac{\alpha_\theta U_\theta}{2}
+ \frac{e^{4U}}{2R^2}\left({A_R}^2 - \alpha {A_\theta}^2\right),
\label{evol2}\\
A_{RR} - \alpha A_{\theta\theta} &=& \frac{A_R}{R}
+ \frac{\alpha_R A_R}{2\alpha} + \frac{\alpha_\theta A_\theta}{2}
- 4A_R U_R + 4\alpha A_\theta U_\theta.
\label{evol4}
\end{eqnarray}
\begin{description}
\item[Auxiliary Equations:]
\end{description}
\begin{eqnarray}
G_R &=& - A H_R
\label{aux1}\\
H_R &=& \frac{e^{2\beta}K}{\sqrt{\alpha} R^3}.
\label{aux2}
\end{eqnarray}
Here $K$ is the (generally nonvanishing) twist constant from
expression~(\ref{nonvanishingtwistconstant}).  Since the change
$K \rightarrow -K$ can be achieved by a coordinate transformation
preserving all the conditions we have imposed,
it is enough to consider $K \geq 0$.

It follows from equations (\ref{constr1}), (\ref{constr3}),
(\ref{aux1}) and (\ref{aux2}) that if we specify smooth initial data
\begin{equation}
{\cal{D}}=\{U, U_R, A, A_R, \alpha, \beta, G, H, K\},
\label{initialdata}
\end{equation}
with $\alpha > 0$ and with $K \geq 0$ a constant,
then the first time derivatives of $\beta, \alpha, G$ and $H$ are
determined in terms of $\cal D$ and its spatial derivatives.  We
then verify that the initial value problem consisting of data $\cal D$
satisfying~(\ref{constr2}) together with the evolution equations
(\ref{evol2}), (\ref{evol4}) and the equations (\ref{constr1}),
(\ref{constr3}), (\ref{aux1}) and (\ref{aux2}) constitutes a well
posed system. Further, as proven in BCIM, we have the following:

\begin{proposition} (BCIM)
\label{BCIMProp}
Let $(\gamma, \pi)$ be a set of smooth $T^2$ symmetric initial data for
the vacuum Einstein equations on $T^3$, with the data such that the 
spacetime gradient of the area of the isometry orbits does not vanish.
For some nonnegative constant $R_0$,
there exists a globally hyperbolic spacetime $(M^4,g)$ such that (a)
$M^4=T^3\times (R_0, \infty)$; (b) $g$ satisfies the vacuum Einstein
equations; (c) $M^4$ is covered by areal coordinates $(x,y,\theta, R)$
with $R\in(R_0,\infty)$, so the metric takes the form (\ref{metric});
and (d) $(M^4,g)$ is isometrically diffeomorphic to the maximal
Cauchy development of the data $(\gamma, \pi)$.
\end{proposition}

This result is effectively a global existence theorem for the areal
coordinate initial value problem (with data $\cal D$) cited above.
In addition, it tells us that, except for the flat 
solutions with group orbits of constant area,
areal coordinates cover the maximal spacetime development of
any vacuum $T^2$ symmetric initial data\footnote{Note that the
conclusion of proposition \ref{BCIMProp}  holds regardless of
whether the initial data is chosen with constant $R$.} on $T^3$.
This result does not, however, tell us for which solutions the
range of the areal time coordinate $R$ extends to all positive
values.  We resolve this question in this paper.  If $K=0$ we
obtain the $T^3$ Gowdy solutions, and for these, $R_0 = 0$
\cite{Moncrief}.  Thus we are interested in the case $K \neq 0$,
though the method of proof applies to the case $K=0$ as well.

In the course of proving proposition \ref{BCIMProp}, BCIM derive some
useful properties of $\beta$, $h$, and the energy-like integral of $h$,
\begin{equation}
{\cal E}(R) = \int_{S^1}h \, d\theta.
\label{calE}
\end{equation}
We now state and prove some of these properties, together with related
results which will be useful in proving our main theorem.

\begin{lemma}
\label{EMono}
Let $R_i\in(R_0,\infty)$.
\begin{enumerate}
\item If ${\cal E}(R_i) =0$ then ${\cal E} (R) = 0$ for all $R\in
(R_0,\infty)$.
\item  If ${\cal E}(R_i) \neq 0$ then ${\cal E}(R)$ is
nonincreasing in $R$ on $(R_0,\infty)$.
\item  If ${\cal E}(R_i) \neq 0$ and $K \neq 0$ then ${\cal E}(R)$
is strictly decreasing in $R$ on $(R_0,\infty)$.
\end{enumerate}
\end{lemma}
$\mathbf {Proof}$: If ${\cal E}(R_i) =0$ then it follows from
(\ref{h}) and (\ref{calE}) that $U_R$, $U_\theta$, $A_R$ and
$A_\theta$ all vanish at $R = R_i$ for all $\theta$.  Therefore,
$U_{\theta \theta}$ and $A_{\theta \theta}$ vanish at $R = R_i$
as well.  But then it follows from the evolution equations that
$U_{RR}$ and $A_{RR}$ vanish for all $\theta$ at $R=R_i$. Since
$U =$ constant and $A=$ constant constitute a solution to the
evolution equations which is  consistent with any data at $R = R_i$
for which ${\cal E}(R_i) =0$, and since the data at $R = R_i$ uniquely
determine a solution for all $R\in (R_0,\infty)$, we necessarily have
${\cal E} = 0$ for all $R$.

Based on the definition of $h$ and using the evolution equations and
the constraints, we readily calculate the time derivative of $h$:
\begin{equation}
h_R = -\frac{e^{2\beta} K^2}{2 R^3} h
-\frac{2 {U_R}^2}{\sqrt{\alpha} R}
-\frac{\sqrt{\alpha}e^{4U} A_\theta^2}{2R^3}
+\frac{(\sqrt{\alpha} \beta_\theta)_\theta}{R}.
\label{h_R}
\end{equation}
If we now integrate both sides of (\ref{h_R}) over the circle, we
have a formula for the time derivative of $\cal E$:
\begin{eqnarray}
{\cal E}_R & = & - \int_{S^1} \left(\frac{e^{2\beta} K^2}{2 R^3} h
+ \frac{2 {U_R}^2}{\sqrt{\alpha} R}
+ \frac{\sqrt{\alpha}e^{4U} A_\theta^2}{2R^3}\right) d\theta.
\label{calEdot}
\end{eqnarray}
Clearly ${\cal E}_R$ is nonpositive. Noting from the above argument
that if ${\cal E}(R_i) \neq 0$ then, for each $R \in (R_0,\infty)$,
$h(\theta,R)\neq 0$ on some open subset of $S^1$, and noting that if
$K \neq 0$ this implies that the first term of (\ref{calEdot}) is
nonvanishing (and negative), we deduce that if ${\cal E}(R_i) \neq 0$
and $K \neq 0$ then ${\cal E}(R)$ is strictly decreasing.
$\Box$

\begin{lemma}
\label{beta-alpha}
\begin{enumerate}
\item
For any $R\in(R_0,\infty)$, the function $\beta$ satisfies the
condition
\begin{equation}
\max_{S_1} \beta(\theta,R) -
\min_{S_1} \beta(\theta,R)\leq R {\cal E} (R).
\label{maxminbeta}
\end{equation}
\item
For any pair $R_1, R_2\in(R_0,\infty)$, the function $\alpha$ satisfies
\end{enumerate}
\begin{equation}
\sqrt{\frac{\alpha(\theta,R_1)}{\alpha(\theta,R_2)}} =
\, e^{2 (\beta(\theta,R_1) - \beta(\theta,R_2))}
\exp\left\{\int^{R_2}_{R_1} 2 \sqrt{\alpha(\theta,R)} \, R \, 
h(\theta,R) \, dR \right\}.
\label{alpha}
\end{equation}
\end{lemma}
$\mathbf {Proof}$: The spatial max-min estimate for $\beta$ follows
from the inequality $Rh\pm\beta_\theta \geq 0$ (or equivalently
$|\beta_\theta| \leq Rh$), which holds because $Rh\pm\beta_\theta$
is the sum of squares (see the constraint equation formula for
$\beta_\theta$ and the definition of $h$). Once we have
$|\beta_\theta| \leq Rh$, we may calculate
(for any $\theta_1, \theta_2$ satisfying $0< \theta_2-\theta_1<1$),
\begin{eqnarray}
|\beta(\theta_2,R) - \beta(\theta_1,R)| & = & 
\left|\int^{\theta_2}_{\theta_1}
\beta_\theta \, d\theta \right|, \nonumber \\
& \leq & \int^{\theta_2}_{\theta_1} |\beta_\theta| \, d\theta,
\nonumber \\
& \leq & \int^{\theta_2}_{\theta_1} R h \, d\theta, \nonumber \\
& \leq & R {\cal E}. \nonumber
\end{eqnarray}
The estimate (\ref{maxminbeta}) immediately  follows.

The result (\ref{alpha}) for $\alpha$ follows directly from
integrating
\begin{eqnarray}
2 \beta_R - \frac{\alpha_R}{2 \alpha} = 2 \sqrt{\alpha} R h
\end{eqnarray}
over time (R).
$\Box$

\section{Analysis for initial data  with ${\cal E} (R_i) \neq 0$}
\label{result}

For a given set of initial data $\cal D$ with $K \neq 0$ (see
(\ref{initialdata})), specified at time $R_i$, the
quantity which determines if the range of the areal time coordinate
(covering the maximal spacetime development of $\cal D$) extends
to all positive values is ${\cal E} (R_i)$.  In this section, we
show that if ${\cal E} (R_i)$ is not zero (whether or not $K = 0$),
then $R_0$ must be zero. (Recall that, as shown in lemma~\ref{EMono},
if ${\cal E} (R_i) \neq 0$, then ${\cal E} (R) \neq 0$ for all
$R\in (R_0,\infty)$.)

So let us assume, as we will throughout this section, that we have
chosen data $\cal D$ at time $R_i$ for which ${\cal E}(R_i) \neq 0$.
We know from proposition \ref{BCIMProp} that the domain of the
maximal Cauchy development of $\cal D$ includes
$S^1\times (R_s,R_i]$ for some positive number\footnote{We reserve
``$R_0 $" to denote the first value of the time -- proceeding from
$R_i$ into the past, towards $0$ -- to which we cannot extend
the domain of dependence of the data $\cal D$. As we shall show
here, if ${\cal E}\neq 0$, we can extend past $R_s$ so long as
$R_s>0$.} $R_s$, with $0 < R_s < R_i$.  In particular, we know that,
for any $R_a \in (R_s, R_i)$, all the metric functions
$\{\alpha, \beta, U,A, G,H\}$ and their derivatives are bounded
on $S^1 \times [R_a,R_i]$, and we know further that $\alpha$ is
bounded away from zero and $\alpha_R$ is nonpositive on this same
region.  What we will now show is that the metric functions, their
first derivatives and the second derivatives of $U$ and $A$ are
bounded and $\alpha$ is bounded away from zero on
$S^1 \times (R_s,R_i]$  so long as $R_s > 0$.  It then follows
that $R_0=0$.

We start by arguing that $\cal E$ extends to $R_s$.

\begin{lemma}
\label{calEbound}
${\cal E}$ is bounded from above on $(R_s,R_i]$ and therefore has a
unique continuous extension to $R_s$.
\end{lemma}
$\mathbf{Proof}$: As discussed in the proof of lemma \ref{EMono}, if
${\cal E}(R_i)\neq 0$, then ${\cal E}(R)$  monotonically decreases into
the future (and therefore monotonically increases into the past). To
control the evolution of ${\cal E}$ into the past, it is useful to
define a closely related quantity,
\begin{equation}
\tilde {\cal E} = {\cal E} + \int_{S^1}
\frac{e^{2 \beta} K^2}{4 \sqrt{\alpha}R^4} \, d\theta,
\label{tildecalE}
\end{equation}
which also evolves monotonically:
\begin{equation}
\tilde {\cal E}_R =
- \int_{S^1} \left(\frac{e^{2\beta}K^2}{\sqrt{\alpha} R^5}
+ \frac{2 {U_R}^2}{\sqrt{\alpha} R}
+ \frac{\sqrt{\alpha}e^{4U} A_\theta^2}{2R^3}\right) d\theta.
\label{tildeEmonotone}
\end{equation}
(The monotonic quantity defined by equation (43) in \cite{ARW} for $T^2$
symmetric solutions of the Einstein-Vlasov equations 
reduces to $\tilde {\cal E}$ in the vacuum case we are considering here.)
Now if we compare the right hand side of (\ref{tildeEmonotone}) with
the definition (\ref{tildecalE}) of $\tilde {\cal E}$ we derive
\begin{equation}
\tilde {\cal E}_R \geq  - \frac{4\tilde {\cal E}}{R}.
\end{equation}
We readily determine that, for any $R_k \in (R_s,R_i]$,
\begin{equation}
\tilde {\cal E}(R_k) \leq \tilde {\cal E}(R_i) + \int_{R_k}^{R_i}
\frac{4\tilde {\cal E}(R)}{R} \, dR.
\end{equation}
Applying Gronwall's inequality lemma to this inequality (as suggested
on page 353 of \cite{a1}), we obtain
\begin{equation}
\tilde {\cal E}(R_k) \leq \tilde {\cal E}(R_i)
\left(\frac{R_i}{R_k}\right)^4,
\end{equation}
for any $R_k \in (R_s,R_i]$.  Thus
$\tilde {\cal E}(R_k) < \tilde {\cal E}(R_i) (R_i/R_s)^4$
on $(R_s,R_i)$.  This bound, together with the
monotonicity of $\tilde {\cal E}(R)$, guarantees that
$\tilde {\cal E}(R)$ continuously extends to $R_s$.

Now it follows from the definition (\ref{tildecalE}) of
$\tilde {\cal E}$ that ${\cal E}(R) \leq \tilde {\cal E}(R)$.
Therefore, noting the monotonicity of ${\cal E} (R)$, we see that
it too extends continuously to $R_s$, thus proving the lemma.
$\Box$

Using the fact that $\cal E$ is defined continuously on $[R_s,R_i]$,
we now argue that the function $\beta$ is bounded from above:

\begin{lemma}
\label{betabound}
$\beta$ is bounded from above on $S^1\times(R_s, R_i]$.
\end{lemma}
$\mathbf{Proof}$: Using the evolution equation (\ref{calEdot}) for
$\cal E$ together with the oscillation result for $\beta$ from
lemma~\ref{beta-alpha}, we calculate the following:
\begin{eqnarray}
{\cal E}(R_s) & \geq & {\cal E}(R_i)
+ \int_{S^1} \left( \int^{R_i}_{R_s}
\frac{e^{2\beta}K^2}{2R^3} h \, dR \right) d\theta, \nonumber \\
& \geq & {\cal E}(R_i) + \int_{S^1} \left( \int^{R_i}_{R_s}
\frac{e^{2 \min_{S^1} \beta}K^2}{2R^3} h \, dR \right) d\theta, \\
& = &  {\cal E}(R_i) + \int^{R_i}_{R_s}
\frac{e^{2 \min_{S^1} \beta}K^2}{2R^3} {\cal E} \, dR, \nonumber \\
& \geq & {\cal E}(R_i) + \int^{R_i}_{R_s}
\frac{e^{2 \max_{S^1} \beta} e^{-2 R \cal E}K^2}{2R^3} {\cal E} \, dR,
\nonumber \\
& \geq & {\cal E}(R_i)\left\{ 1  + e^{-2 R_i {\cal E}(R_s)}
\int^{R_i}_{R_s}
\frac{e^{2 \max_{S^1} \beta}K^2}{2R^3} \, dR\right\}. \nonumber
\end{eqnarray}
Therefore, for any $\theta \in S^1$, we have
\begin{equation}
{\cal E}(R_s) \geq
{\cal E}(R_i)\left\{ 1  + e^{-2 R_i {\cal E}(R_s)} \int^{R_i}_{R_s}
\frac{e^{2 \beta(\theta,R)}K^2}{2R^3} \, dR\right\}.
\label{Eunbounded}
\end{equation}
Assuming that ${\cal E}(R_i)\neq 0$, this gives us the integral
inequality
\begin{equation}
\int^{R_i}_{R_s} \frac{e^{2 \beta(\theta,R)}K^2}{2R^3} \, dR
\leq e^{2 R_i {\cal E}(R_s)}
\left(\frac{{\cal E}(R_s)}{{\cal E}(R_i)} - 1 \right). \nonumber
\end{equation}
If we now integrate the time evolution equation (\ref{constr1}) for
$\beta$, and use the inequality which we just derived, then for each
$R_k \in (R_s, R_i]$ and for each $\theta \in S^1$, we obtain
\begin{eqnarray}
\beta(\theta,R_k) & = & \beta(\theta,R_i) +
\int^{R_i}_{R_k} \frac{e^{2 \beta(\theta,R)}K^2}{4R^3} \, dR
- \int^{R_i}_{R_k} \sqrt{\alpha} R h \, dR, \nonumber \\
& \leq & \beta(\theta,R_i) + \int^{R_i}_{R_k}
\frac{e^{2 \beta(\theta,R)}K^2}{4R^3} \, dR, \nonumber \\
& \leq & \beta(\theta,R_i) + 
\frac{1}{2}e^{2 R_i {\cal E}(R_s)}
\left(\frac{{\cal E}(R_s)}{{\cal E}(R_i)} - 1 \right), \nonumber\\
& \leq & \max_{S^1} \beta(\theta,R_i)+
\frac{1}{2}e^{2 R_i {\cal E}(R_s)}
\left(\frac{{\cal E}(R_s)}{{\cal E}(R_i)} - 1 \right).
\label{calEboundsbeta}
\end{eqnarray}
This inequality determines an upper bound for
$\beta$ on $S^1 \times (R_s,R_i]$, thereby proving the lemma.
$\Box$

We next argue that $\alpha$ is controlled:

\begin{lemma}
\label{alphabound}
$\alpha$ is bounded from above and away from zero on
$S^1\times(R_s, R_i]$, and has a unique continuous extension to
$S^1\times \{R_s\}$.  Also, the time derivative $\alpha_R$ is
bounded above and below on $S^1\times(R_s, R_i]$.
\end{lemma}
$\mathbf{Proof}$: Using the evolution equation (\ref{constr2})
for $\alpha$, we derive an equation for $(\ln \alpha)_R$ which
is nonpositive and only involves functions which are bounded
on $S^1 \times (R_s, R_i]$. Integrating over time, we see that
$\ln \alpha$ is bounded on this region.  Since it is
monotone in $R$ it extends to $S^1\times \{R_s\}$.
The stated results for $\alpha$ and $\alpha_R$ then follow.
$\Box$

With control of $\alpha$ established and $\beta$ bounded from above,
we can now use $1+1$ light cone arguments to show that the rest of
the metric functions are bounded and that $\beta$ is bounded from
below on $S^1\times (R_s,R_i]$.  We start with the functions $U$
and $A$ and their first derivatives:

\begin{lemma}
\label{lightcones}
The functions $U$, $A$ and their first derivatives are
bounded on $S^1\times(R_s, R_i]$.
\end{lemma}
$\mathbf{Proof}$: We follow the pattern of the light cone arguments
in sections 4 and 6 of BCIM. So let us first define
\begin{eqnarray}
E &=& \frac{\sqrt{\alpha} h}{2}, \nonumber \\
&=& \frac{U_R^2}{2}+\frac{\alpha U_\theta^2}{2}
+\frac{e^{4U}}{8R^2} \left(A_R^2 + \alpha A_\theta^2 \right),
\label{E} \\
P &=& \frac{\sqrt{\alpha} \beta_x}{2R}, \nonumber \\
&=& \sqrt{\alpha}
\left(U_R U_\theta+\frac{e^{4U}}{4R^2} A_R A_\theta \right).
\label{P}
\end{eqnarray}
The derivatives of the sums and differences of these two quantities
along the two null directions
\begin{eqnarray}
\partial_l &=& \partial_R+\sqrt{\alpha}\partial_\theta, \\
\partial_n &=& \partial_R-\sqrt{\alpha}\partial_\theta.
\label{nullvectors}
\end{eqnarray}
are readily calculated. We obtain
\begin{eqnarray}
\partial_n(E+P) &=& J_{+}, \\
\partial_l(E-P) &=& J_{-},
\label{E&Peqns}
\end{eqnarray}
with $J_+$ and $J_-$ defined by
\begin{equation}
J_\pm = -\frac{e^{2\beta} K^2}{R^3}(E \pm P)
- \frac{U_R^2}{R}-\frac{\alpha e^{4U} A_\theta^2}{4R^3}
\mp \frac{P}{R}.
\label{J}
\end{equation}
Note that since $\alpha$ extends to $S^1\times [R_s,R_i]$, the vector
fields $\partial_l$ and $\partial_n$ do as well.

Before proceeding with the integration of $E$ and $P$ along null paths
generated by $\partial_l$ and $\partial_n$, we note some useful
inequalities which follow immediately from the definitions of $E$,
$P$, $J_+$ and $J_-$ which we have just given. First, since $E \pm P$
are both the sum of squares, we have
\begin{equation}
|P|\leq E.
\end{equation}
Then, as a consequence of this result together with the definitions,
we have
\begin{equation}
|J_\pm| \leq \left(\frac{2 e^{2\beta} K^2}{R^3} + \frac{3}{R}\right) E.
\end{equation}

We now consider an arbitrary point
$(\theta,R_k) \in S^1\times (R_s,R_i)$.  We let $\gamma_l$ and
$\gamma_n$ denote the null paths (generated by $\partial_l$ and
$\partial_n$, respectively) which meet at $(\theta,R_k)$, and we
let $p_l$ and $p_n$ denote the intersection of each of these paths
with the initial data surface $S^1\times\{R_i\}$. If we integrate
the evolution equations (\ref{E&Peqns}) along $\gamma_l$ and
$\gamma_n$, combine the results, and use some of the estimates
established above, we have
\begin{eqnarray}
E(\theta,R_k) &=& \frac{1}{2}\left\{E(p_l)- P(p_l) + E(p_n) + P(p_n)
- \int_{\gamma_l} J_- - \int_{\gamma_n} J_+ \right\}, \nonumber \\
&\leq& E(p_l) + E(p_n) \\
&& \; \; + \int_{\gamma_l}
\left(\frac{e^{2\beta} K^2}{R^3} + \frac{3}{2R}\right) E
+ \int_{\gamma_n} \left(\frac{e^{2\beta} K^2}{R^3} +
\frac{3}{2R} \right) E. \nonumber
\end{eqnarray}
It follows from this inequality, and from consideration of
maxima over $S^1$, that we have
\begin{eqnarray}
\max_{S^1} E(\theta,R_k) & \leq & 2  \max_{S^1} E(\theta,R_i) \\
&& \; \; + \int_{R_k}^{R_i}
\left(\frac{2 e^{2 \max_{S^1} \beta(\theta,R)} K^2}{R^3} +
\frac{3}{R}\right) \max_{S^1}E(\theta,R) \, dR. \nonumber
\end{eqnarray}
Then applying the 
Gronwall inequality lemma to this result, we obtain
$$\max_{S^1} E(\theta,R_k) \leq 2 \max_{S^1}E(\theta,R_i)
\exp\left\{\int_{R_k}^{R_i}
\left(\frac{2e^{2 \max_{S^1} \beta(\theta,R)} K^2}{R^3} +
\frac{3}{R}\right) dR \right\}.$$
From this it follows that
\begin{equation}
E(\theta,R_k) < 2 \max_{S^1}E(\theta,R_i)
\exp\left\{\int_{R_s}^{R_i}
\left(\frac{2e^{2 \max_{S^1} \beta(\theta,R)} K^2}{R^3} +
\frac{3}{R}\right) dR \right\},
\label{boundE}
\end{equation}
which determines an upper bound on $S^1 \times (R_s,R_i]$
for the positive quantity $E(\theta,R)$.

Since $E$ and $\ln \alpha$ are bounded, it follows that $h$ is as
well.  Then, it follows from the definition of $h$ that both $U_R$
and $U_\theta$ are bounded.  Integrating along appropriate paths in
$S^1 \times (R_s,R_i]$, we obtain a bound on $U$.  Then, since $R$
is bounded away from zero, it follows from the definition of $h$ that
both $A_R$ and $A_\theta$ are bounded.  Integrating along appropriate
paths, we obtain a bound on $A$.
$\Box$

We have thus far determined that
$U, U_R, U_\theta, A, A_R, A_\theta, \ln \alpha$ and $\alpha_R$
are bounded on the region of interest, $S^1 \times (R_s,R_i]$.
We need to show that the same is true for $\beta$, for the rest
of the first derivatives of $\alpha$ and $\beta$, and
also for the second derivatives of $U$ and $A$. We obtain some of
these results by direct calculation from the field equations (see
lemma \ref{alphabetaderivs}).  For the others, we use a light
cone argument, as sketched in lemma \ref{2ndderivs}.

\begin{lemma}
\label{alphabetaderivs}
The functions $\beta$, $\alpha_\theta$, $\beta_\theta$ and $\beta_R$
are bounded on $S^1 \times (R_s,R_i]$.
\end{lemma}
$\mathbf{Proof}$: Since $U, A$ and their first derivatives are
bounded and since $\beta$ is bounded from above, it follows from the
formulas (\ref{constr1})-(\ref{constr2}) for $\beta_R$ and
$\beta_\theta$ that these two functions are also bounded.
Thus $\beta$ is bounded from below as well as from above.

To bound $\alpha_\theta$, we work indirectly: We first take the
spatial derivative of equation (\ref{constr3}), obtaining
\begin{equation}
\alpha_{R\theta} = \frac{\alpha_R \alpha_\theta}{\alpha}
+ 2 \alpha_R \beta_\theta.
\label{getalphax}
\end{equation}
We then integrate this equation over time, which gives us
(for any $R_k\in(R_s, R_i]$)
\begin{eqnarray}
\alpha_\theta(\theta, R_k) & = & \exp\left\{-\int_{R_k}^{R_i}
\frac{\alpha_R}{\alpha} \, dR \right\}
\Bigg[ \alpha_\theta(\theta,R_i) \nonumber \\
&& \hspace{40pt} - \int_{R_k}^{R_i} 2 \alpha_R \beta_\theta
\exp\left\{\int_R^{R_i}
\frac{\alpha_s}{\alpha} \, ds \right\} dR \Bigg].
\label{alphax}
\end{eqnarray}
The right hand side of equation~(\ref{alphax}) is bounded, so
$\alpha_\theta$ must be bounded.
$\Box$

\begin{lemma}
\label{2ndderivs}
The second derivatives of $U$ and $A$ are bounded on
$S^1 \times (R_s,R_i]$.
\end{lemma}
$\mathbf{Proof}$: Just as the control of the first derivatives of $U$
and $A$ is obtained via a light cone argument based on the quantities
$E$ and $P$ defined in equations (\ref{E})-(\ref{P}), we control the
second derivatives of $U$ and $A$ using similar, but higher order,
quantities. These are defined as follows:
\begin{eqnarray}
E_1 & =& \frac{1}{2} \left\{ U_{RR}^2 + \alpha \, U_{R\theta}^2
+ \frac{e^{4U}}{4 R^2}(A_{RR}^2
+ \alpha A_{R\theta}^2) \right\}, \\
P_1&=&\sqrt{\alpha} \left( U_{RR}U_{R\theta}
+\frac{e^{4U}}{4 R^2} A_{RR} A_{R\theta}\right).
\end{eqnarray}
Carrying out an argument similar to that used in
lemma~\ref{lightcones}, we calculate the null derivatives
of sums and differences of $E_1$ and $P_1$:
\begin{eqnarray}
\partial_n(E_1+P_1)&=&L_{+},
\label{E1eqn} \\
\partial_l(E_1-P_1)&=&L_{-}.
\label{P1eqn}
\end{eqnarray}
Here $L_{+}$ and $L_{-}$ are expressions which we shall not
display explicitly.  The key feature of them which we need,
that on $S^1\times (R_s, R_i]$ we have
\begin{equation}
|L_\pm| \leq f_1(\theta,R) + f_2(\theta,R) E_1,
\end{equation}
with $f_1$ and $f_2$ bounded, is readily derived.
We then integrate equations (\ref{E1eqn})-(\ref{P1eqn}) along
null paths generated by $\partial_l$ and $\partial_n$, and derive
estimates for $E_1$ which guarantee that it is bounded.
Boundedness for $U_{RR}, U_{R\theta},  A_{RR}$, and $A_{R\theta}$ then
follows.  To show that $U_{\theta\theta}$ and $A_{\theta\theta}$ are
bounded, we rely on the evolution equations (\ref{evol2}) and
(\ref{evol4}), in which all terms except for  $U_{\theta\theta}$ and
$A_{\theta\theta}$ have already been determined  to be bounded.
$\Box$

All that remains is to control the shift functions $G$ and $H$:

\begin{lemma}
\label{auxil}
$G$ and $H$ and their first derivatives are bounded
on $S^1 \times (R_s,R_i]$.
\end{lemma}
$\mathbf{Proof}$: The time derivatives of the shift functions $G$ and
$H$ are specified by equations (\ref{aux1}) and (\ref{aux2}). Noting
that the right hand sides of these equations consist entirely of
bounded functions, we see that $G_R$ and $H_R$ are
bounded and we integrate $G_R$ and $H_R$ along appropriate
paths in $S^1 \times (R_s,R_i]$ to obtain bounds on $G$ and $H$.
Similarly, differentiating equations (\ref{aux1}) and (\ref{aux2})
with respect to $\theta$ and then integrating with respect to
time, we obtain bounds on $G_\theta$ and $H_\theta$.
$\Box$

We may now prove the basic result of this section:

\begin{proposition}
\label{extension}
Let $\cal D$ be data at $R_i > 0$ for a vacuum $T^2$ symmetric
solution of the Einstein equations with ${\cal E}\neq 0$. The
maximal Cauchy development of $\cal D$ is covered by areal coordinates
with the areal time coordinate R taking on all positive real values.
\end{proposition}
$\mathbf{Proof}$: Suppose that some development of $\cal D$ with
${\cal E}\neq 0$ is covered by areal coordinates with
$(\theta,R) \in S^1 \times (R_s, R_i]$, for some $R_s$ such that
$0<R_s<R_i$.  As a consequence of lemmas~\ref{calEbound} through
\ref{auxil}, we then know that all of the functions $U$, $A$, $\beta$,
$\ln \alpha$, $G$ and $H$ are bounded on $S^1\times (R_s, R_i]$, and
their first derivatives and the second derivatives of $U$ and $A$
are bounded on this spacetime region as well.  With local existence
established by proposition 1, then standard long-time existence
theorems for PDEs (see, for example, theorems 2.1 and 2.2, as well
as corollaries~1 and 2 in\footnote{The
system of first order equations for
$(\beta, \alpha, U, U_R/\sqrt{\alpha}, U_\theta,
A_R/\sqrt{\alpha}, A_\theta)$
satisfies the hypotheses of these theorems
and corollaries.} chapter 2 of Majda's book \cite{Majda})
show that the solution in terms of the areal coordinate metric
functions $U$, $A$, $\beta$, $\alpha$, $G$ and $H$ extends to
$S^1 \times (0,R_i)$.  Proposition~\ref{BCIMProp} then shows that
the maximal Cauchy development of $\cal D$ is covered by areal
coordinates with the areal time coordinate $R$ taking on all
positive real values.
$\Box$

\section{Analysis for initial data  with ${\cal E} (R_i) = 0$}
\label{flatKasner}

In section 3, we have shown that if a set of data $\cal D$ has
${\cal E}\neq 0$ then the maximal development of $\cal D$ has
$R_0=0$. In this section, we consider what happens if the data
$\cal D$ has ${\cal E}= 0$ and $K \neq 0$. We first show that the
areal coordinates covering the maximal development of $\cal D$
have $R_0>0$. We then show that ${\cal E} =0$ if and only if the
maximal development of the data is flat Kasner.

\begin{proposition}
\label{E=0R>0}
Let $\cal D$ be a set of smooth initial data for a $T^2$ symmetric
solution, with ${\cal E}=0$ and $K \neq 0$.  The areal coordinates
for the maximal development of $\cal D$ have $R_0>0$.
\end{proposition}
$\mathbf{Proof}$: If ${\cal E}=0$, then it follows from the definition
of $\cal E$ and from lemma \ref{EMono} that $h=0$ for all $\theta$
and $R$. Examining equations (\ref {constr1}) and (\ref {constr2}),
we then see that $\beta_\theta = 0$ (so $\beta$ is independent of
$\theta$) and that the evolution of $\beta$ is governed by
\begin{equation}
\beta_R = -\frac{e^{2\beta}K^2}{4R^3}.
\label{evbeta}
\end{equation}
Since K is constant, this is a separable ODE, which we can readily
integrate to get (presuming that the initial data is specified at
time $R=R_i$)
\begin{equation}
e^{- 2 \beta(\theta,R)} = e^{- 2 \beta(\theta,R_i)}
+ \frac{K^2}{4R_i^2}-\frac{K^2}{4R^2}.
\label{beta/R}
\end{equation}

We notice that, as we evolve in areal coordinates back in time,
starting at $R=R_i$, the right hand side of equation (\ref{beta/R})
becomes smaller. Finally, for $R$ equal to some positive number
$R_0<R_i$ , the right hand side of equation (\ref{beta/R}) goes to
zero. It follows from equation (\ref{beta/R}) that as $R$ approaches
this value $R_0$, $\beta$ grows without bound.  Since
proposition~\ref{BCIMProp} tells us that areal coordinates cover the
maximal development of the data $\cal D$, and since $\beta$ blowing
up signals the end of areal coordinates, we conclude that the maximal
development of $\cal D$ ends with $R$ equal to this value of $R_0$,
defined by
\begin{equation}
0 = e^{- 2 \beta(\theta,R_i)} + \frac{K^2}{4R_i^2}-\frac{K^2}{4R_0^2}.
\label{R_0}
\end{equation}

Note that, since $\beta$ is independent of $\theta$, $R_0$ is as well.
$\Box$

\begin{proposition}
\label{flatkasner}
The development of a set of data $\cal D$ with $K \neq 0$ is isometric
to flat Kasner if and only if ${\cal E}=0$.
\end{proposition}
$\mathbf{Proof}$: First, let us assume that we are given a set of data
with ${\cal E}=0$. It follows from the definition of ${\cal E}=0$ that
$U$ and $A$ must be constant in both $\theta$ and $R$. Examining the
constraint equation (\ref{constr2}), we see that $\beta$ must be
constant in $\theta$ as well.  The functions $\alpha$, $G$ and
$H$ are not necessarily independent of $\theta$, but spatial
coordinate transformations that preserve the conditions we have
imposed can be used to remove any $\theta$-dependence of these
functions.  The presence of three nowhere vanishing, spatial,
linearly independent, commuting Killing vectors and calculation
of the Kasner exponents shows that this is flat Kasner.

To argue the converse, let us assume that the $T^2$ symmetric spacetime
$\{M^4, g\}$ we are considering is $T^3$ flat Kasner.  Then we may
choose a fixed set of three nowhere vanishing, linearly independent,
commuting Killing fields $\{W_i\}$ tangent to the homogeneous Cauchy
surfaces. It follows
that the Killing fields $X$ and $Y$ which identify the spacetime as
$T^2$ symmetric are each linear combinations of the $\{W_i\}$ with
constant coefficients.\footnote{Note that we are considering general
compactifications of the flat Kasner spacetime, with no assumption
being made that the integral curves of $\{W_i\}$ or of $X$ or of
$Y$ are closed.}  It is also true, by assumption, that the twist
quantity of expression (\ref{vanishingtwistconstant}) vanishes,
while that of expression (\ref{nonvanishingtwistconstant}) does not.
If we now calculate the range of $R$, defined to be the area of the
orbits generated by $X$ and $Y$, on the maximal Cauchy development
of any one of the $R=$ constant surfaces, we find that it is
$(R_0,\infty)$ with $R_0 > 0$.  (This is true regardless of our
choice of $\{W_i\}$.) From proposition~\ref{extension} it follows
that ${\cal E} = 0$.
$\Box$

We note that, combining propositions~\ref{extension} and \ref{E=0R>0},
we see that for vacuum $T^2$ symmetric spacetimes with nonvanishing
twist constant, $R_0=0$ if and only if ${\cal E}\neq0$.  We should
emphasize that this result presumes that the twist quantity $K$ of
(\ref{nonvanishingtwistconstant}) is
nonzero and that the twist quantity (\ref{vanishingtwistconstant})
vanishes.  If both twist quantities vanish ({\it i.e.}, for Gowdy
spacetimes), then the  areal time coordinate covers the full range
$(0,\infty)$, regardless of whether ${\cal E}$ is zero or not (and
regardless of whether the spacetime is flat Kasner or not).  If the
twist quantity (\ref{vanishingtwistconstant}) does not vanish,  then
${\cal E}(R_i)\neq0$ does not imply that $R_0 = 0$.  However, if
the spacetime is flat Kasner then $R_0 > 0$, while  if the spacetime
is not flat Kasner then $R_0 = 0$.

\section{Conclusion}
\label{conclusion}

Combining the results of sections 3 and 4, we have proven our main
theorem.

\begin{theorem}
\label{main}
On the maximal Cauchy development of vacuum $T^2$ symmetric initial
data for the Einstein equations with nonvanishing twist constant,  the
area of the $T^2$ group orbits takes on all positive values if and
only if the solution is not flat Kasner. Hence the areal coordinates
$(x,y,\theta,R) \in T^3 \times (R_0, \infty)$ which cover every such
spacetime have $R_0=0$ if and only if the spacetime is not flat Kasner.
\end{theorem}

This result should be a useful tool for the study of these spacetimes.
It tells us that in terms of areal coordinates, one approaches the
initial cosmological singularity exactly by letting $R$ approach zero.

There is considerable interest in analyzing the nature of the
singularities in the vacuum $T^2$ symmetric spacetimes. Numerical
studies combined with heuristic argument strongly suggest that,
generically, the cosmological singularity in these spacetimes is
oscillatory ({\it i.e.}, ``mixmaster'') in nature \cite{BIW}.   This
is true for other classes of spacetimes as well ({\it e.g.}, those
with $U(1)$ symmetry \cite{BM98}).  However, there has never been
a proof of generic oscillatory behavior in an infinite
dimensional\footnote{See \cite{BM00} for a finite dimensional
class of spatially inhomogeneous oscillatory solutions.}
class of spatially inhomogeneous
solutions, and the $T^2$ symmetric class of spacetimes is among the
simplest such class in which oscillatory behavior is expected. So
there is considerable incentive to study the singularities in these
spacetimes further, and our result should be helpful for such studies. 

We note that heuristic analyses suggest that $R_0=0$ is a prerequisite
for oscillatory behavior to occur, and such analyses coupled with
numerical work do indicate (in agreement with our results here)
that generally $R_0=0$ \cite{Weav2}. We also note that it has
been shown (in \cite{ARW}) that in $T^2$ symmetric solutions,
$R=R_0$ is a crushing singularity. This means that a constant
mean curvature foliation exists near the singularity, with the
mean curvature growing without bound. 

Our result here has an interesting corollary concerning the
existence of solutions which are flat Kasner in some regions
but not elsewhere. In $T^3$ Gowdy spacetimes, this can happen
near the singularity. Specifically, in $T^3$ Gowdy, if initial
data is given at $0< R_i < 1/2$, and if the data is flat Kasner
on some interval $[a,b]$, with $b - a > 2 R_i$, then the region
of the spacetime development of this data which is locally
isometric to flat Kasner will extend to the singularity.  For
$T^2$ symmetric solutions with  $K \neq 0$, however, our result
shows that even if $R_i > 0$ is very small and even if the
proper subset of the initial $S^1$ on which the data is flat
Kasner is very large, so long as there is some point of the
initial $S^1$ at which $h \neq 0$ (and therefore, as a
consequence of continuity, there is an interval on which
$h \neq 0$) then there is no region abutting the singularity
in which the spacetime is locally isometric to flat Kasner.
This corollary rules out, for $T^2$ symmetric solutions with
$K \neq 0$, a particular kind of construction of solutions with
Cauchy horizons which can be carried out for Gowdy spacetimes.

\section{Acknowledgments}
This work was partially supported by the National Science Foundation
under grant PHY-0099373 at Oregon, and by the Natural Sciences and
Engineering Research Council of Canada.  Part of this work
was carried out at the Workshop on General Relativity,
April 1 - June 7, 2002, Stanford Mathematics Department and
American Institute of Mathematics.  MW also thanks the Institute
of Theoretical Science at the University of Oregon for hospitality,
November-December, 2001, at the beginning stages of this work.

\end{document}